\documentclass{camera}
\usepackage{graphicx}  
\usepackage{natbib}

\begin{document}

%
\title{RELATIVISTIC SHOCKS AND ULTRA HIGH ENERGY COSMIC RAY ORIGIN}

%
\author{Jacek Niemiec}

%
\organization{Instytut Fizyki J\c{a}drowej PAN, ul. Radzikowskiego 152, 31-342
Krak\'ow Poland }

\maketitle

\begin{abstract}
The current status of the theory of particle acceleration at relativistic shocks
is presented, and a few possible scenarios for ultra-high-energy cosmic 
ray production at such shocks are briefly discussed.
\end{abstract}

%



%

\section{Introduction}
Relativistic shock waves are widely thought to generate energetic particles
producing high-energy emission in many astrophysical sources. The example of
such sources can be hot spots in radio galaxies, jets in blazars and in Galactic
``microquasars", gamma-ray burst sources, and pulsar winds. It is also believed
that high-energy cosmic rays have their origin in astrophysical environments in
which relativistic shocks occur. The aim of this paper is to present the current
status of cosmic ray acceleration theory at relativistic shock waves and to
briefly review and comment on the models proposed for ultra-high-energy cosmic 
ray production at such shocks.  

\section{First-order Fermi Process at Relativistic Shocks}
The basic acceleration mechanism discussed in the context of cosmic ray 
production at shocks is the first-order Fermi process. This process can be
applied to sufficiently energetic particles, with gyroradii much larger
than the shock thickness, defined by gyroradii of ``thermal''
ions present in the plasma. Such energetic particles 
see the shock front as a sharp discontinuity in the plasma flow.
In the following, we review the results obtained mainly in the test particle 
approach, in which the influence of accelerated particles on the physical
conditions at the shock is not taken into account. 
 
In the first-order Fermi acceleration process particles gain their energies by
consecutive crossings of the shock front. In order to do so, they have to be
confined near the shock. The particle confinement is provided by the turbulent
magnetic fields which perturb particle trajectories leading to particle 
diffusion in pitch-angle. During the acceleration process some fraction of 
particles transmitted downstream escapes from
the vicinity of the shock. The competition between particle energy gains at
shock interactions and the escape process determines the stationary particle 
spectrum, often of the power-low form.

In the case of a nonrelativistic shock, where the fluid speeds are much lower 
than the energetic particle velocity, the resulting particle spectrum is 
independent of the 
background conditions near the shock, including the configuration of the
regular magnetic field, the spectrum and the amplitude of MHD turbulence. 
This is mainly because of a nearly isotropic form of the particle distribution 
function at the shock, in the conditions where  magnetic
field perturbations near the shock are sufficient to ensure efficient 
particle scattering. In such conditions, the spectral index $\alpha$ for the 
phase-space
distribution function is given exclusively by the shock compression ratio $R$, 
and $\alpha = 3R/(R-1)$, 
where $R=u_1/u_2$ is given in the shock normal rest frame, and $u_1$ and $u_2$ 
are the shock velocities with respect to the upstream and downstream plasma rest
frames, respectively. 

The physical picture is much more complicated for relativistic shocks, where the
shock velocity or its projection along the upstream mean magnetic field 
$u_{B,1}=u_1/\cos\psi_1$ ($\psi_1$ -- the upstream magnetic field inclination 
angle to the shock normal) is comparable to the speed of light. 
The energetic particle distribution becomes 
anisotropic near the shock, and anisotropy increases with growing shock 
Lorentz factor $\gamma$. This fact substantially influences the resulting particle 
spectrum, which is then very sensitive to the background conditions. 

Studies of particle acceleration processes at mildly relativistic shock waves
started with the works of \citet{kir87a} and \citet{hea88} for parallel shocks 
($\psi_1 \equiv 0^o$). 
They showed that the spectral indices are
different from the $\alpha=4$ value obtained in the nonrelativistic case, and
depend on the form of the wave power spectrum of the magnetic field
perturbations \citep[see also][]{ell90}. Acceleration processes in
oblique subluminal ($u_{B,1}<c$) shock waves were analyzed by 
\citet{kir89} under the assumption of magnetic moment conservation for particles
interacting with the shock. This assumption restricted validity of their
considerations to the case of a weakly perturbed magnetic field, where 
cross-field diffusion does not play a significant role. They showed
that in such conditions particle spectra can be even as flat as $\alpha\approx
3$ in cases where $u_{B,1}$ is close to the speed of light. This feature results
from effective multiple reflections of anisotropically distributed upstream
particles from the compressed field downstream of the shock. 

\begin{figure}[t]   
\vspace{6cm} 
\includegraphics{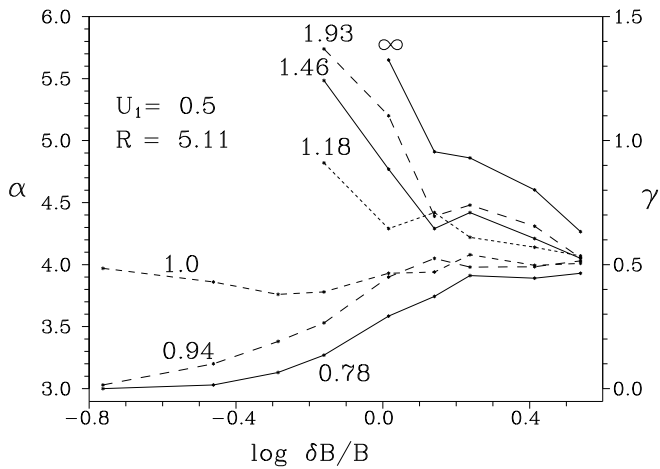}    
\vspace{0.4cm} 
\caption{Spectral indices for the oblique shock waves moving with velocity 
$u_1=0.5c$ versus perturbation amplitude $\delta B/B$ for different
inclinations of the mean magnetic field, given in terms of $u_{B,1}$ near the 
respective curves \citep{ost02b}.
}
\end{figure}

The feature discussed is also visible in fig. 1, which shows  
results of Monte Carlo particle simulations performed by
 \citet{ost93}. In the figure, one may note the lack of results for 
superluminal ($u_{B,1}>c$) shocks propagating in a weakly perturbed magnetic
field. In such conditions, particles are tied to the field lines.
Therefore, the upstream particles can be only transmitted downstream with no
possibility of returning to the shock, so the first-order Fermi process does not
operate in this case and the power-law particle spectrum cannot be formed.
Nevertheless, because of anisotropy of particle distribution at a relativistic
shock, particle energy gains in a transmission process can be much larger than
those resulting from the adiabatic compression at the shock \citep{beg90}. 

In the effectively accelerating astrophysical shocks one may expect
large-amplitude MHD waves to occur. The role of such finite-amplitude magnetic
field perturbations in forming a particle spectrum was investigated by a~number
of authors using Monte Carlo simulations \citep[e.g.,][]{ell90,
ost91,ost93,bal92,bed96,bed98}. The derived power-law particle spectra 
directly depend on the conditions near the shock. They can be either very steep
or very flat for different configurations of the mean magnetic field with
respect to the shock normal and different amplitudes of perturbations, as
shown in fig. 1. Note also, that the changes of the particle spectral
index can be nonmonotonic with an increasing field perturbations amplitude
\citep{ost91,ost93}. In the case of a highly perturbed magnetic field, power-law
spectra can be formed for superluminal shocks. They are, however, very 
steep for mildly perturbed conditions at the shock.

The first-order Fermi 
process at ultrarelativistic shocks has been recently 
discussed  by a number of authors \citep[e.g.,][]{bed98,gal99,ach01,kir00}.
This topic is of a great importance for gamma-ray burst external shocks
astrophysics and UHECR generation in these sources.  
For $\gamma\gg 1$
the energetic particle distribution
function is highly anisotropic at the shock in the upstream plasma rest 
frame. This stems from the fact that when a given particle crosses the shock 
upstream, its momentum direction has an opening angle around the shock normal 
$\sim 1/\gamma$,  thus it is almost aligned with the shock propagation direction. 
In such conditions, a small perturbation to the particle trajectory, provided
by its momentum deflection in the upstream regular magnetic field or scattering 
off the MHD fluctuations, allows for its being overtaken by the shock and 
transmitted downstream. As a result, the mean particle energy gain in a single 
acceleration loop upstream-downstream-upstream is $\Delta E/E \sim 1$
\citep{bed98,ach01}, which restricts
seriously the efficiency of the acceleration process.  

Efficient particle acceleration is possible when particle distribution upstream
is close to isotropy. However, such conditions can be met at the 
ultrarelativistic shock wave only at the particle first shock crossing. Then,
the mean energy gain for particles reflected from the shock is very large,
$\Delta E/E \sim \gamma^2$,
but it should be stressed that the expected efficiency of such
reflection is rather low \citep{ach01}. As explained above, the conditions at 
the shock do not allow for particle distribution isotropization in 
subsequent shock crossings, even if a highly perturbed
medium exists upstream of the shock.

The important topic in the context of UHECR production at high-$\gamma$
shock waves is the existence of the asymptotic spectral index for relativistic
shock acceleration. All the above
mentioned works on the first-order Fermi process at such shocks yield consistent
estimates of the accelerated particle spectral index $\alpha\approx 4.2-4.3$, 
in the limit of the very high shock Lorentz factors ($\gamma\rightarrow\infty$).
The models of the burst afterglow spectra often give results pointing to 
essentially the same value of $\alpha$ for synchrotron radiating electrons,
which is sometimes interpreted as observational confirmation of 
the correctness of theoretical models proposed for ultrarelativistic shock 
acceleration. However, in all discussed derivations there is explicit or 
implicit assumption of a highly turbulent conditions near the shock
\citep[see][]{ostb02}. It is uncertain whether such conditions are indeed met in
astrophysical shocks. On the other hand, in conditions with medium-amplitude
magnetic field perturbations, particle spectra generated at oblique 
realistic shocks
with $\gamma\sim 10-100$ can be much steeper than those obtained in the
asymptotic limit, as seen in the simulations of \citet{bed98}
\citep[see also][]{nie04}. In such conditions, the shocks will 
be unable to efficiently accelerate particles in the first-order Fermi process. 

The factors important for 
the first-order Fermi processes at relativistic shocks are the presence and the
configuration of the mean magnetic field, the structure of the turbulent field
component, and also the presence of the long-wave magnetic field perturbations.
The acceleration studies presented above apply very simple approaches for the
modeling of the turbulent MHD medium near the shock. Their perturbed magnetic
field models either lack some of the factors mentioned above or take them into 
account in a very simplified way. In some models, the continuity of the 
perturbed field across the shock is also not preserved. This factor 
is, however, very important for the acceleration process since it can 
introduce the correlations in particle motion on both sides of the shock.
There is thus a need for a more detailed analysis of the first-order Fermi 
process at relativistic shocks, which would include all the important factors 
to yield the better insight into the real physical situation at the 
astrophysical shocks.

An attempt at doing such analysis for mildly relativistic shocks has been 
recently performed by \citet{nie04}. We considered the
first-order Fermi process at shocks propagating in ``more realistically'' 
modeled perturbed magnetic fields. The model assumes a wide
wavevector range turbulent field component with the power-law spectrum imposed 
on the uniform magnetic field component upstream of the shock. The continuity of
the full perturbed field across the shock is preserved by the use of the 
respective shock jump conditions. In our modeling, the particle spectra were 
derived with the method of Monte Carlo simulations by integrating particle 
trajectories in such magnetic field, with a hybrid method enabling to treat both
long- and short-wave perturbations.

\begin{figure}[t]   
\includegraphics[scale=0.75]{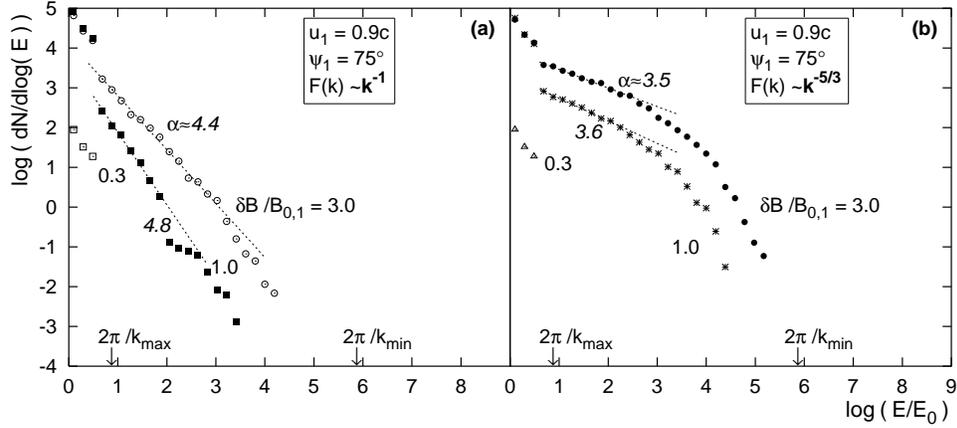}
\vspace{-0.65cm} 
\caption{Particle spectra at the oblique superluminal shock with $u_1=0.9c$ for
(a) the flat and (b) the Kolmogorov wave spectrum of the magnetic
field turbulence. The upstream perturbation amplitude $\delta B/B_{0,1}$ is given
near the respective curves. Linear fits to the spectra are also presented and
values of the spectral indices $\alpha$ are given. The spectra have vertical
shifts for clarity. Particles with energies in the range 
$(2\pi/k_{max}, 2\pi/k_{min})$ can effectively interact with the field
perturbations.}
\end{figure}

Figure 2 summarizes the results for oblique superluminal shocks, where the
particle spectra for the shock with  
$u_1=0.9c$ $(u_{B,1}\simeq 3.48)$  
are shown as an example.
This is a representative example for
relativistic shocks, because most field configurations in such shocks lead to
analogous superluminal conditions \citep{beg90}.
One can note that particle spectra diverge from a simple power law in the full
energy range. For the small turbulence amplitude the particle spectrum is
essentially the 
compressed upstream injected distribution.
 Power-law
sections in the spectra are produced at larger turbulence amplitudes, but in our
simulations they
are formed by a small number of particles. The spectra steepen with growing
energy, but still within the resonance range, and are followed by the cutoffs. 

These features arise due
to the finite wavevector range of the turbulence considered in the simulations. 
In the perturbed magnetic field with the limited dynamic range of field
perturbations, scattering conditions vary with particle energy. Low-energy
particles have their resonance wavevectors close to the maximum wavevector value
$k_{max}$, so that most of the turbulence reside in the long-wave range for this
particles. These long-wave, high-amplitude magnetic field perturbations form
{\it locally} oblique, subluminal field configurations at the shock, which enable 
acceleration of particles interacting with the shock in the 
locally subluminal conditions \citep[see also][]{ost93}. 
The acceleration process becomes less efficient for high-energy particles due to
decreasing amount of the respective long waves for these particles, which
results in the spectrum
steepening and the cutoff. This feature is independent of the initial energy of
particles injected at the shock \citep[sec. 3.3]{nie04}, suggesting that
actually the turbulent magnetic field structure does not allow for particle
acceleration to higher energies.

From the above considerations it is evident that the critical component of
any shock acceleration study is the applied model of the perturbed magnetic
field near the shock. The real magnetic field structure and its generation 
mechanism at a relativistic shock front are poorly known. Therefore, an
understanding of these and other related issues (e.g., particle injection,
the shock structure) is the basic condition of reaching the real progress in the
study of cosmic ray acceleration at relativistic shock waves.

A noticeable advance in this field results from the application of
particle-in-cell (PIC) simulations, developed to study
``microphysical''
properties of relativistic shocks \citep{dru01,sch02,nis03,sil03,fre04}.
As the theoretical considerations show \citep{med99}, the turbulent magnetic 
field can be generated locally at the shock via relativistic two-stream
(Weibel) instability. The source and the mechanism of the field generation 
is kinetic in nature, as the instability is driven by the anisotropy of particle
distribution function at the shock associated with a two-stream motion of the
two plasma particle populations: the first formed by the upstream particles
inflowing onto the shock front and the second composed of particles reflected
from the shock. The more detailed insight into the dynamics of the Weibel 
instability give us particle-in-cell simulations, in which two relativistic
plasma collide with each
other to form a shock. The structure of the small-scale strong turbulent 
magnetic field, generated downstream of the shock, is very complicated spatially
and, in addition, changes with time. The field is predominantly 
transversal and the total magnetic energy 
becomes a substantial fraction of the equipartition field energy. 
The initial perturbations grow in size and the generated field
structures are transported downstream. At the given point downstream of the
shock and a given time long after collision, the power spectrum of the turbulent 
magnetic field is a power-law, with the most power concentrated at long 
wavelengths. The scales associated with the magnetic fields 
thus generated are, however, small compared to gyroradii of particles 
undergoing the first-order Fermi acceleration. Nonetheless, the magnetic field 
structure evolution is probably accompanied by plasma heating and non-thermal 
particle acceleration, 
thus possibly providing the seed particles for the 
Fermi process. The PIC simulations can be therefore treated as a first step 
towards understanding of processes important for the study of cosmic ray 
acceleration at relativistic shocks.

In summary, the present theoretical knowledge of cosmic ray production at
relativistic shocks is incomplete. 
Existing theoretical models explain only some
features of the observed cosmic rays. Because of the strong dependence of
particle spectra on conditions at the shock, the models do not allow for
reliable modeling of astrophysical objects in which relativistic shock waves
occur. Further progress requires an increase in the number and the quality of 
observational data revealing the physical parameters near the shock. 
On the other hand, there is a need for advance in theoretical studies,
which most probably is to be made through advanced numerical simulations.
Particle-in-cell methods are appropriate to study the microphysics of
relativistic shocks. Realistic modeling of particle acceleration at
astrophysical shocks has to incorporate results of such studies and by itself 
requires a full plasma nonlinear description. This should take into account 
appropriate boundary conditions, modification of the shock structure by 
backreaction of accelerated particles \citep[see][]{ell02} and second-order 
acceleration processes.  
                                               
\section{UHECRs from Relativistic Shocks}
Extragalactic astrophysical objects harboring relativistic shock waves are
the likely sources of ultra-high-energy cosmic rays, with energies in the range
$E\sim 10^{18}-3\cdot 10^{20}$ eV. In the light of the apparent drawbacks in the
theory of particle acceleration at relativistic shocks, it is clear that it 
is now vary hard to judge with certainty which mechanism proposed for UHECR 
production is in fact responsible for the observational data on cosmic rays. 
Nevertheless, it is worthy to comment on some proposals on this topic.

\subsection{Gamma-ray Burst Internal and External Shocks}
Fireball models of gamma-ray bursts assume the prompt burst emission in gamma
rays to be produced at internal mildly relativistic shocks ($\gamma \sim 2-10$)
and the afterglow emission to originate at the blast wave propagating with
initially
ultrarelativistic velocity ($\gamma \sim 10^2-10^3$) into the external medium
\cite[see, e.g.,][]{zha03}. The evidence for ultra-high-energy cosmic ray
production at GRB sources is  based on two claimed coincidences
\citep{wax04}. First, the constraints imposed on fireball model parameters,
inferred from prompt gamma-ray and early afterglow observations, are similar to
the constraints imposed on the shock in order to allow for proton acceleration
to the energies above $10^{20}$ eV. Second, energy generation rate in gamma rays
by gamma-ray bursts is similar to the rate required to account for the 
UHECR energy density. The model of \cite{wax04} assumes
cosmic ray acceleration to very high energies by the first-order Fermi processes
\citep[see also][]{wax95} at internal shocks. The claimed agreement of model
estimates with the observed UHECR flux and spectrum is however based on an 
oversimplified assumption about the spectrum of accelerated
protons, which is taken as the one derived for non-relativistic shocks, where
$\alpha=4$. Such assumption cannot be justified for mildly relativistic
shocks, where spectral indices are predicted to be much steeper in the most
probably realized cases of superluminal shock configuration. Moreover, the
medium into which an internal shock propagates may be already modified, either by
previous shell collisions, leading to a highly relativistic temperature of the
upstream plasma, or by magnetic field amplification processes. In the latter
case, the close to equipartition magnetic fields may occur near the
shock, so that the magnetic field may play a non-negligible dynamical
role. Both effects lead to the steeper particle spectra than those
obtained at unmodified shocks \citep{kir00}. Note also, that in conditions of a
highly turbulent close-to-equipartition fields near the relativistic shock the
{\it second-order} acceleration processes may play a significant role in UHECR
production \citep{der01}, modifying the resulting particle
spectrum.

The proposal of UHECRs production at GRB 
external ultrarelativistic shocks, based on a claimed evidence of the
first-order Fermi acceleration seen in burst afterglow data (see sec. 2),
was first advanced by \citet{vie95}. Analysis performed by
\citet{gal99} shows however, that the Fermi process is 
inefficient in accelerating particles 
when the blast wave propagates in the Galactic-like interstellar medium. 
The maximum energy a particle can gain in this case is
$E_{max}\sim 5\cdot 10^{15} Z B_{\mu G}$ eV. Therefore, in order to reach 
higher particle energies, the acceleration process must operate in the region 
with much stronger magnetic fields.   
Note, however, that the generation of high-energy
particles at ultrarelativistic shocks probably requires the presence of 
long-wave field perturbations  \citep[see][]{nie04}, even 
if a highly turbulent medium exists near the shock. 
 
Ultra-high particle energies can be also reached at GRB external shocks 
if there are energetic particles 
preexisting in the medium into which a blast wave propagates. These particles,
when reflected from the shock, examine the $\gamma^2$ energy boost. 
The particle reflection process is efficient if the energy density content in
relativistic particles $\epsilon_R$ is sufficiently large, as the fraction of 
the blast wave energy that can go into boosting these particles to UHECR 
energies is proportional to $\epsilon_R/\epsilon_{tot}$. 
Cosmic rays  
of energy above $\sim 10^{14}$ eV (sufficient to reach $10^{20}$ eV if $\gamma 
> 10^3$) in the Galactic-like interstellar medium
comprise only a very small fraction $\sim 10^{-9}$ of the total ISM energy 
density, and the reflection process cannot account for the observed UHECRs, 
as indicated by  \citet{gal99}. More promising scenario is
when the fireball blast wave expands into the pulsar wind bubble. Then, the
bubble energy density is dominated by energetic ions, which can be boosted to
UHECR energies with high efficiency. As shown in \citep{gal99}, 
spectra of the boosted ions are determined by the blast wave deceleration, and
the evaluated spectral index is $\alpha=4$ in this case. This leads to the 
UHECR spectrum 
compatible with observations. Note, however, that predictions of the presented 
model are based on a number of assumptions
about physical conditions which lead to the most efficient acceleration. 
To be considered seriously as the source of UHECRs, this mechanism should be 
modeled for definite realistic conditions, including the magnetic field 
structure in the pulsar wind bubble. 

\subsection{Hot Spots of FRII Radio Sources}
The possible sites for UHECR production are also the hot spots of FRII radio 
sources. These are believed to be the downstream regions of mildly relativistic
shocks formed by interaction of relativistic jets with intergalactic medium.
The model of cosmic ray production at hot spots 
was presented by 
\citet{rac93},
who assumed that protons are accelerated in 
the first-order Fermi process at
non-relativistic parallel shock. From comparison of the acceleration and the
radiative loss time scales, they estimated $E_{max}$ to lie in the UHECR range.
They show that in order to explain the observed UHECR flux, the particle spectral
index at the shock must be on average $\alpha =4$ or slightly flatter. Although
this is in agreement with recent results obtained on cosmic ray 
acceleration at mildly relativistic parallel shocks \citep{nie04} (results for
subluminal shocks may also apply given the typical hot spot shock speeds 
$u_1=0.3c$), the assumption
made by Rachen \& Biermann of the Kolmogorov magnetic field turbulence
downstream, extending up to the scales corresponding to gyroradii of the most
energetic protons, may not be valid (see sec. 2). Moreover, it is not
certain how much the finite size of the hot spots, and corresponding diffusive 
particle losses, influence the accelerated proton
spectrum. Authors' estimates of the critical energy above which these losses
lead to the spectrum cutoff give 
(for a hot spot size $L\sim 1$ kpc and the magnetic field $B\simeq 0.5$ mG) 
$E_{c\parallel} \approx 10^{18}$ eV for diffusion parallel to
the mean magnetic field and $E_{c\perp} \approx 4\cdot 10^{19}$ eV for
cross-field diffusion. These estimates depend on the acceleration time scale,
which may be smaller for oblique shocks \citep{bed96}, but even if the 
perpendicular field
configuration is assumed, the critical energy cannot be much larger then
$E_{c\perp}\sim LBu_2/c \approx 4\cdot 10^{19}$ eV \citep[see][]{ost02}. Note
also, that in that case the proton spectrum is probably very steep. 

\vspace*{-0.2cm}
\subsection{Non-standard Fermi Process at Ultrarelativistic Shocks}
\vspace*{-0.1cm}
It is worthy to mention about an interesting non-standard Fermi mechanism of 
particle
acceleration proposed recently by \citet{der03}. This mechanism
takes advantage of multiple particle conversions (photon-induced or resulting
from nucleon collisions) from
the charged state into neutral state and back. The charged particle (proton)
downstream from the shock, when converted into neutral state (neutron), can
get into the upstream region with no influence from the magnetic field. There,
conversion to the charge state may occur and the particle is magnetically confined
to the plasma flow. At the moment of conversion, the particle can be far from the
shock, so it may be deflected by a large angle before reaching the shock again. 
The distribution of such particles is then isotropic upstream, what
enables them to increase their energies in a whole cycle by a factor of
$\sim\gamma^2$. Because of this, the process can be very efficient in
producing UHECR particles, even if the efficiency of the converter mechanism 
alone is low. It should however be investigated in more details, 
also using numerical simulations.  
  
\vspace*{-0.2cm}
\section{Final Remarks}
\vspace*{-0.1cm}
The purpose of this work was to present the current understanding of the
first-order Fermi acceleration processes at relativistic shocks. The present 
knowledge in that matter is confronted with various proposals of UHECR production at
sources harboring relativistic shocks. It was not the aim of the work to fully
discuss these models, we comment only on the features related to the assumed 
details of acceleration processes. There are however additional conditions that
should be
met by the models to account for the observed UHECR characteristics. Any
successful model, apart from being able to fit the spectrum of UHECRs, has also 
to match their composition and
arrival direction distribution, account for possible small-scale clustering of
events etc. Such the issues are described in detail in, e.g.,
 \citep{oli00}.

\section*{Acknowledgements}
I am grateful to Micha\l{} Ostrowski for enlightening discussions and critical
remarks on the manuscript. The work was supported by the Polish State 
Committee for Scientific Research through the grant
PBZ-KBN-054/P03/2001.

\end{document}